\documentclass[twocolumn]{article}
\usepackage{graphicx}
\begin{document}

\begin{center}
{\bf Dynamics of spin correlations\\
in the spin--$\frac{1}{2}$ isotropic $XY$ chain\\ 
in a transverse field}\\
\vspace{1mm}
O. Derzhko$^\dagger$, 
T. Krokhmalskii$^\dagger$, 
and J. Stolze$^\ddagger$\\
\vspace{1mm}
$^\dagger$Institute for Condensed Matter Physics\\
1 Svientsitskii Str., L'viv--11, 79011, Ukraine\\
derzhko@icmp.lviv.ua
\hspace{2mm}
krokhm@icmp.lviv.ua\\
\vspace{1mm}
$^\ddagger$Institut f\"ur Physik, Universit\"at Dortmund\\
44221 Dortmund, Germany\\
stolze@physik.uni-dortmund.de
\end{center}

One of the simplest quantum spin systems 
is the spin--$\frac{1}{2}$ $XY$ chain 
introduced by Lieb {\it et al} [1], 
who pointed out the relation between the spin model 
and noninteracting spinless fermions 
that makes many properties of the spin model 
amenable to analytic calculations. 
Dynamic spin correlations are of particular interest 
since they can be measured 
by neutron scattering or magnetic resonance techniques 
in a number of materials 
which are reasonably well described by different 
special cases of the spin--$\frac{1}{2}$ $XY$ chain 
(e.g., 
CsH$_2$PO$_4$, PbHPO$_4$, PrCl$_3$, CsCuCl$_3$ 
etc). 
Unfortunately, 
the calculation of dynamic spin correlations 
can be nontrivial 
even in the simple case of the $XY$ chain, 
and the complete 
wavevector and frequency dependences 
of the dynamic structure factors 
or the dynamic susceptibilities 
have been obtained only recently 
using a numerical approach [2].
In what follows 
we present the results 
for the time--dependent two--spin correlation functions 
of the spin--$\frac{1}{2}$ isotropic $XY$ chain in a transverse field 
obtained numerically 
and apply these findings 
to discuss the temperature dependence 
of the spin--spin relaxation time in PrCl$_3$ [3] 
which was not known heretofore 
due to lack of information 
about the correlation functions 
for the spin--$\frac{1}{2}$ isotropic $XY$ chain. 

We consider $N$ spins one--half 
governed by the following Hamiltonian
$$
H=\Omega\sum_{n=1}^N
+J\sum_{n=1}^{N-1}\left(s_n^xs_{n+1}^x+s_n^ys_{n+1}^y\right)
$$ 
where $\Omega$ is the transverse field 
and $J$ is the exchange interaction. 
Using the numerical approach explained in detail in Ref. [2] 
we compute the time--dependent two--spin correlation functions
$\langle s_j^{\alpha}(t)s_{j+n}^{\beta}\rangle$,
$\alpha\beta=xx,xy,zz$.
Here the angular brackets stand for 
the canonical thermodynamic average 
with the inverse temperature $\beta=1/k_BT$.
In figure A 
\begin{figure}[h]
\includegraphics[width=3in,clip]{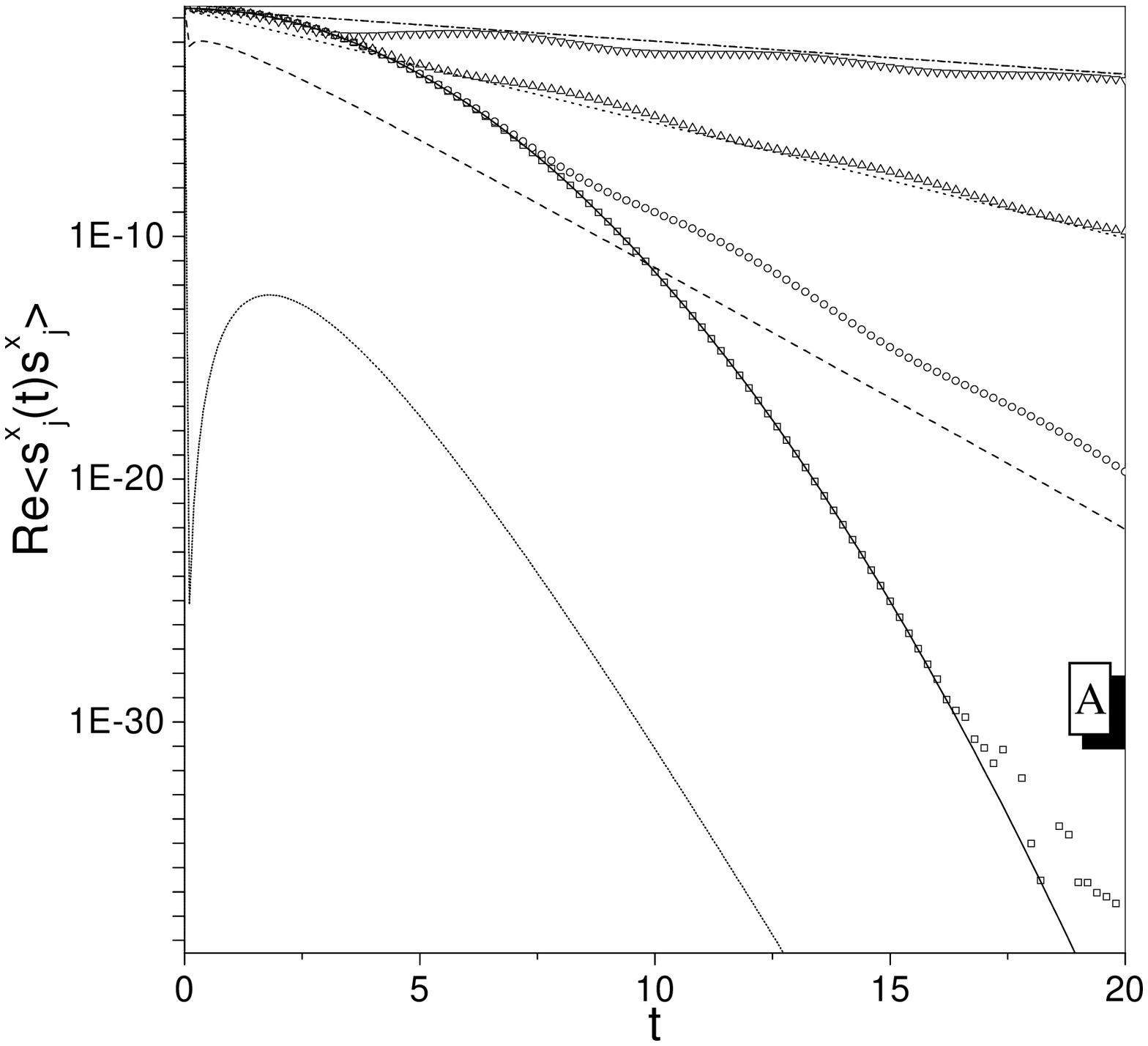}
\end{figure}
we show the numerical results 
for the time dependence of the real part 
of the $xx$ autocorrelation function for $\Omega=0$, $J=-1$ 
at different temperatures. 
In our numerical computations we take $N=400$, $j=51$.
Downward and upward triangles, circles, and squares correspond to
$\beta=5,\;1,\;0.1,$ and $0.00001$, respectively.
We also plot 
the exact analytical result for infinite temperature $\beta=0$ [4]
by a solid curve. 
Dot--dashed, short--dashed, long--dashed, and dotted curves 
correspond to the asymptotic result [5] 
for $\beta=5,\;1,\;0.1,$ and $0.00001$, respectively. 
From figure A 
one can see an excellent agreement 
between the numerical and analytical results 
(only the slopes of asymptotics should be compared with numerical results).
At finite temperatures 
$\langle s_j^x(t) s_j^x\rangle$ decays exponentially 
(the logarithms of the asymptotic results 
have the same slopes 
as the logarithms of the computed correlation functions);
with increasing temperature 
one observes Gaussian decay over a time interval of increasing length
(the plotted data for $\beta=0.00001$ 
show that the correlation function at this temperature 
does not reach its asymptotic regime 
within the time range displayed in the figure A).

The time--dependent two--spin correlation functions 
yield the dynamic structure factors or the dynamic susceptibilities 
that can be measured in materials
which are reasonably well described 
by the spin--$\frac{1}{2}$ $XY$ chain.
As an application of our study 
on the time dependence of the spin correlations 
we reconsider the theoretical prediction 
for the temperature dependence 
of the spin--spin relaxation time in PrCl$_3$ [3].
The Pr--Pr interaction in this compound 
as derived from measurements of static quantities 
can be reasonably well described by the model considered  here
with the exchange interaction 
$J/k_B=2.85$ K
and
$\Omega=0$.
The spin--spin relaxation time $T_2$ 
is related to the $xx$ time--dependent autocorrelation function 
$$
\frac{1}{T_2}
\sim \int_{-\infty}^{\infty}{\mbox{d}}t
\langle s_j^x(t) s_j^x\rangle.
$$
In figure B
\begin{figure}
\includegraphics[width=3in,clip]{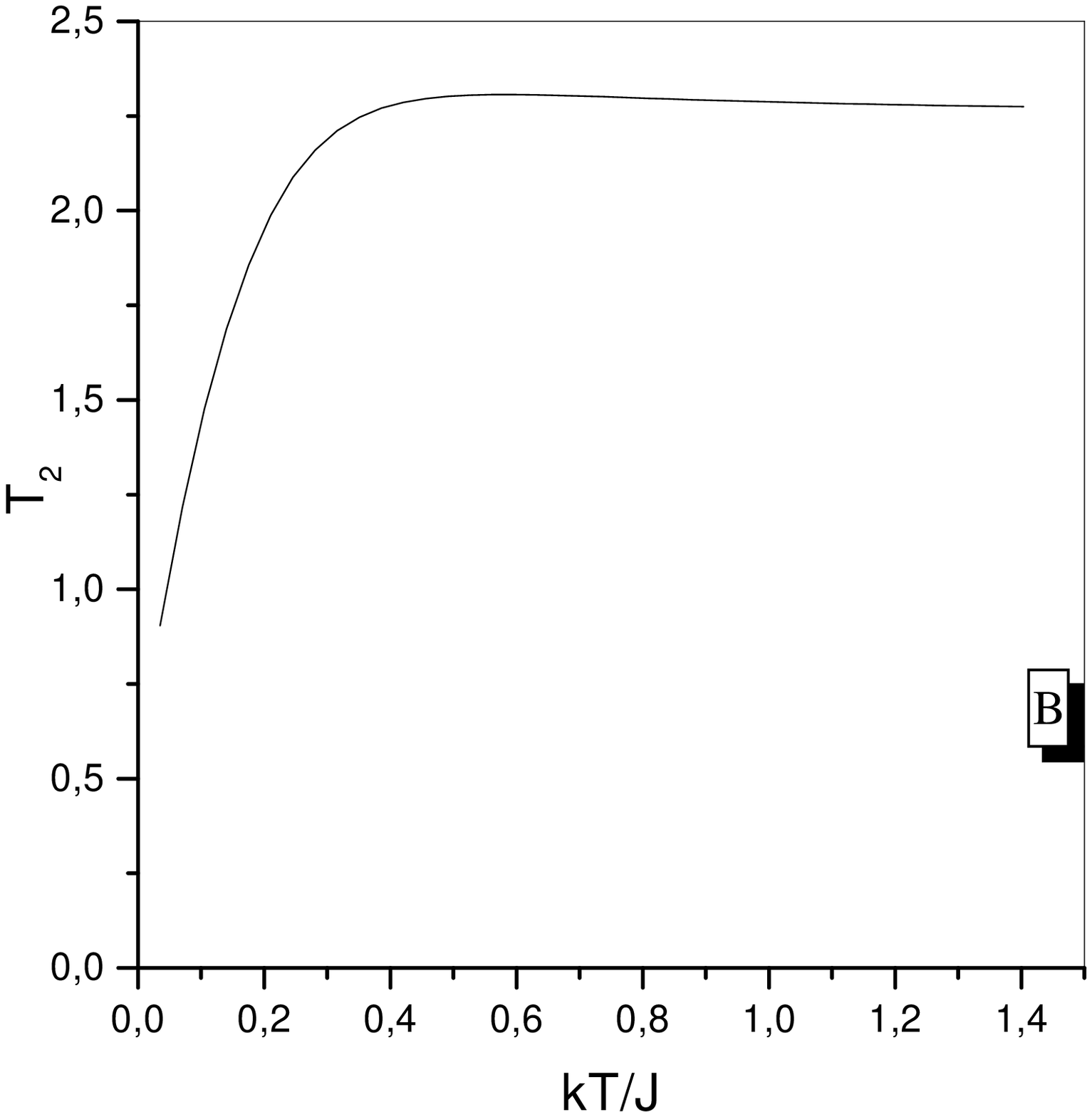}
\end{figure}
we show the result 
of our long--chain calculation 
of the dependence of $T_2$ on temperature.
$T_2$ is an increasing function of temperature 
and crosses over to a constant at about 1 K. 
The plateau in the temperature dependence of $T_2$ 
can be understood from the results presented in figure A.
For the time range during which 
$\langle s_j^x(t) s_j^x\rangle$
contributes appreciably to the integral, 
it does not depend on temperature for sufficiently high temperatures. 
As one can see in figure A 
the data for $\beta\le 1$ 
coincide in the region 
where $\langle s_j^x(t) s_j^x\rangle$ has non--negligible values;
the differences between high and infinite temperatures 
occur only at long times,
where
$\langle s_j^x(t) s_j^x\rangle$ is already very small.

Qualitatively, our result for the temperature dependence 
is similar to that of an earlier short--chain calculation [3]
and to the behaviour observed experimentally [3].
However, there are significant discrepancies 
in the value of the crossover temperature. 
The short--chain calculations yields a crossover at less than 2.5 K;
additional approximations
(coarse--graining in the frequency  $\omega$)
were necessary to extrapolate from $N=10,\;12$ to $N=\infty$ [3].
The experimental crossover temperature of 6 K is still much higher.
This indicates 
(as already suspected earlier [3]) 
that while the static thermal behaviour of PrCl$_3$ 
may be describable by the isotropic $XY$ chain 
with $J/k_B=2.85$ K, 
the dynamics is probably more complicated.

\end{document}